\documentclass[12pt]{iopart}
\usepackage{iopams}
\usepackage{graphicx}
\usepackage{bm}
\usepackage{color}

\begin{document}

\newcommand{\beq}{\begin{equation}}
\newcommand{\eeq}{\end{equation}}
\newcommand{\barr}{\begin{eqnarray}}
\newcommand{\earr}{\end{eqnarray}}

\def\bra#1{\langle{#1}|}
\def\ket#1{|{#1}\rangle}
\def\sinc{\mathop{\text{sinc}}\nolimits}
\def\cV{\mathcal{V}}
\def\cH{\mathcal{H}}
\def\cT{\mathcal{T}}
\renewcommand{\Re}{\mathop{\text{Re}}\nolimits}

\definecolor{dgreen}{rgb}{0,0.5,0}
\newcommand{\green}{\color{dgreen}}
\newcommand{\RED}[1]{{\color{red}#1}}
\newcommand{\BLUE}[1]{{\color{blue}#1}}
\newcommand{\GREEN}[1]{{\color{dgreen}#1}}
\newcommand{\REV}[1]{{\color{red}[[#1]]}}
\newcommand{\KY}[1]{\textbf{\color{red}[[#1]]}}
\newcommand{\SP}[1]{{\color{blue} [[#1 Saverio]]}}
\newcommand{\rev}[1]{{\color{red}[[#1]]}}

\def\HN#1{{\color{magenta}#1}}
\def\DEL#1{{\color{yellow}#1}}

\title{Quantum Typicality and Initial Conditions}

\author{Paolo Facchi$^1$, Saverio Pascazio$^1$, Francesco V.\ Pepe$^1$}

\address{$^1$Dipartimento di Fisica and MECENAS, Universit\`a di Bari, I-70126 Bari, Italy \\
and INFN, Sezione di Bari, I-70126 Bari, Italy}

\begin{abstract}
If the state of a quantum system is sampled out of a suitable
ensemble, the measurement of some observables will yield (almost)
always the same result. This leads us to the notion of quantum
typicality: for some quantities the initial conditions are
immaterial. We discuss this problem in the framework of
Bose-Einstein condensates.
\end{abstract}




\vspace*{1cm}
\noindent
\textbf{A Birthday Dedication}

\noindent Margarita and Volodya Man'ko are a remarkable example of
a life-long passion for physics. Their involvement in fundamental
physics, from quantum optics to quantum mechanics, conveys the
enthusiasm of two teenagers. We are therefore delighted to
dedicate this article to their joint 150th anniversary. Many happy
returns!

\section{Introduction and Motivations}
\label{sec-intro}

When we endeavour to describe the motion of a classical system,
such as a point particle, we write Newton's
equation and a given set of initial conditions. Mathematically, we
try and solve a Cauchy problem for a differential equation. This
is the inheritance of Pierre Simon (Marquis de) Laplace, a world
that is governed by deterministic laws.

A given state of Laplace's deterministic universe is unmistakably
the cause of its future (and the effect of its past). A ``demon"
who at a certain moment knows all forces, positions and velocities
of all particles, would be able to describe their motion with
arbitrary accuracy and a single equation. The future would be
certain to him and he would be able to calculate it from the laws
of classical mechanics. In his work, Laplace never used the word
``demon", which came only later, possibly to convey a feeling of
awkwardness. He rather wrote of ``une intelligence", and for such
an intellect ``nothing would be uncertain and the future just like
the past would be present before its eyes."\footnote{``Rien ne
serait incertain pour elle, et l'avenir comme le pass\'e, seraient
pr\'esent \`a ses yeux." Many of Laplace books are freely
available online, thanks also to an Internet Archive with funding
from the University of Ottawa,
http://www.uottawa.ca/articles/uottawa-library-s-french-language-collection-going-digital}
Laplace was very keen of his deterministic framework. When
Napoleon asked him why he had not mentioned God in his book on
astronomy., he allegedly replied that he ``had no need of that
hypothesis."\footnote{``Je n'avais pas besoin de cette
hypoth\`ese-l\`a".}

Later studies, in particular by Henri Poincar\'e, showed that
Laplace's idea of determinism requires attention and a very
careful scrutiny. The motion of some systems is extremely
``sensitive to the initial conditions" and this has come to be
called dynamical instability. Physicists like the concept of
stability, that makes it meaningful to speak of state preparation,
and guarantees that if one is careful in preparing the state of
the (classical) system, any experiment will yield the same result.
This is known as repeatability and is a milestone of Galileo's
modern scientific method \cite{Russell,Popper,Kline}. Nowadays, to
most physicists, dynamical instability is the same in meaning as
chaos \cite{classchaos}.

Quantum mechanics brought uncertainty (and with it mystery) back
to the stage. Even if one sets the initial conditions of the
Schr\"odinger equation with accurate (infinite) precision, the
behaviour of the (quantum) particle is far from being
deterministic, and is in fact subject to indeterminacy. Quantum
indeterminacy is ontological and not epistemic like in classical
statistical mechanics: it cannot be avoided even by the most
accurate definition of the initial conditions (state preparation).

But is this the whole story? Can one prepare the \emph{very same}
quantum state over and over? This is a very difficult question,
that has mind-boggling aspects.\footnote{In fact, we are convinced
that Margarita and Volodya would like this question!} On one hand,
it is almost meaningless to state that, say, two electrons emitted
by an electron gun and illuminating a double slit have the ``same"
wave function. On the other hand, the experimental verification of this statement,
as e.g.\ through quantum state tomography \cite{Olga97,MarmoPhysScr},
requires measurements over a huge number of
(``identically prepared") electrons. A more cautious question
would then be the following one: which measurements would yield
the same result for a quantum state that is sampled out of a
suitable ensemble? This question catapults us into the topic of
this article and the notion of \emph{quantum typicality}.

It should be clear from the previous discussion that in order to
test the notion of quantum typicality (namely the independence of
the measurement outcome on state preparation), one needs
essentially two ingredients. First, some control is required on
the system state: namely, one must be able to assert, with
reasonable confidence, that the wave function belongs to a
suitable ensemble, e.g.\ a given subspace of the total Hilbert
space of the system, and thus the quantum state is described by a certain density matrix.
This relaxes the very notion of state
preparation: it is not necessary to require (and believe) that a
given wave function is \emph{identically} re-prepared at each
experimental run. It is enough that the wave functions in different runs
are drawn from a suitable statistical ensemble. Second, and equally important, one cannot
expect that measurements yield (almost) the same result for
\emph{any} observable of the system. Some observable will be
typical, other will (and can) not. This is the essence of quantum
mechanics. If one were able to suppress all fluctuations
(including quantum fluctuations) of all observables, the system
would be classical.

It emerges that cold gases and Bose-Einstein condensates (BECs)
are an ideal testbed for these ideas. Indeed, a
BEC of is characterized by a macroscopic occupation of the
\emph{same} single-particle state, or few orthogonal states
(fragmented BEC)~\cite{PS,PeSm}, and one can reasonably assume
that when an experiment is repeated, almost the same wave function
is re-prepared.

These ideas can be tested in double-slit experiments with BECs,
where interference is observed in single experimental runs, even
though the two interfering modes are independently prepared (and
therefore there is no phase coherence)  \cite{exptBEC}. The
presence of an interference pattern is an interesting example of a
property that weakly depends on the choice of the system state:
second-order (unlike first-order) interference is similar for
number and phase states \cite{PS,PeSm} and this explains why
interference patterns emerge in single experimental runs
\cite{JY,CGNZ,WCW,CD,Leggett,Raz,BDZ,YI,AYI,typbec}.

This article is organized as follows. In Section
\ref{sec-typicality} we introduce the statistical ensemble of
quantum states and define the typicality of a quantum observable.
In Section \ref{sec-casestudy} we look at a simple case-study, a
two-mode system of $N$ Bose particles. Section \ref{sec-concl} is
devoted to conclusions and perspectives.

\section{Quantum Typicality}
\label{sec-typicality}

Let a quantum system live in an $N$-dimensional Hilbert space
$\mathcal{H}_N$ and assume that state preparation consists in
\emph{randomly} picking a given pure state $\ket{\Phi_N}$ out of
an $n$-dimensional subspace $\mathcal{H}_n \subset \mathcal{H}$.
Given a basis $\{\ket{\ell}\}$ of $\mathcal{H}_n$, one can write
\begin{equation}
\label{stato}
\ket{\Phi_N} = \sum z_{\ell} \ket{\ell} .
\end{equation}
The complex coefficients $\{z_\ell\}$ are assumed to be uniformly
sampled on the surface of the unit sphere $\sum_\ell
|z_{\ell}|^2=1$.\footnote{This is the simplifying assumption of
\emph{uniform} sampling. Our results are qualitatively unchanged
for a wide class of probability distributions on $\mathcal{H}_n$.}
Clearly
\begin{equation}
\label{distrcoeff}
\overline{z_\ell} = 0, \qquad
\overline{z_{\ell_1}^* z^{ }_{\ell_2}} = \frac{1}{n}
\delta_{\ell_1,\ell_2},
\end{equation}
where the bar denotes the statistical average over the
distribution of the coefficients. Notice the dependence on the
inverse of the subspace dimension $n$ and observe how the average
of all phase-dependent quantities (including the coefficients)
vanish.

Consider an observable $\hat{A}$. The random features of state
(\ref{stato}) will induce fluctuations on a number of quantities
related to $\hat{A}$. We now scrutinize the different origins of
these fluctuations.

The expectation value of observable $\hat{A}$ over state
(\ref{stato}) reads
\begin{equation}
\label{AA}
A= \bra{\Phi_N} \hat{A} \ket{\Phi_N}
\end{equation}
and is itself a random variable. A relevant quantity is the
statistical average of the quantum expectation (\ref{AA}) over the
statistical distribution (\ref{distrcoeff}) of the coefficients
\begin{eqnarray}
\label{expGEN} \overline{A} & := & \overline{ \bra{\Phi_N} \hat{A}
\ket{\Phi_N}} = \tr (\overline{ \ket{\Phi_N} \bra{\Phi_N}}\hat{A}
) \nonumber \\ & = & \frac{1}{n} \sum_{\ell} \bra{\ell} \hat{A}
\ket{\ell} = \tr (\rho_n \hat{A} ) .
\end{eqnarray}
Interestingly, this coincides with the quantum average over the
(totally mixed) ``micro canonical'' density matrix $\rho_n$, which is proportional to
the projector $\hat{P}_n$ onto the subspace $\mathcal{H}_n$:
\begin{eqnarray}
\label{rhomix} \rho_n & = & \overline{\ket{\Phi_N}\bra{\Phi_N}} =
\sum_{\ell_1, \ell_2} \overline{z_{\ell_1} z^*_{\ell_2}}
\ket{\ell_1} \bra{\ell_2} \nonumber \\ & = & \frac{1}{n}
\sum_{\ell} \ket{\ell} \bra{\ell} = \frac{1}{n} \hat{P}_n .
\end{eqnarray}

Another interesting quantity is the statistical variance of the
quantum expectation~(\ref{AA})
\begin{equation}
\label{1}
\delta_s A^2 := \overline{ A^2 } - \overline{A}^2 = \overline{
\bra{\Phi_N} \hat{A} \ket{\Phi_N}^2} - \overline{ \bra{\Phi_N}
\hat{A} \ket{\Phi_N}}^2 .
\end{equation}
If $A$ were deterministic, i.e.\ $\delta_s A \simeq 0$, the
overwhelming majority of states in the statistical ensemble would
have the same expectation value, and this would coincide with the
average $\overline{A}$: this condition defines the {\it typicality
of the expectation value}. Observe that the latter term in
(\ref{1}) involves a quadratic (easy-to-evaluate) average, while
the former term is quartic: since the theory is not Gaussian its evaluation requires some care~\cite{cumulants}.
However, as we shall see, its evaluation is not necessary for our purposes.

A third relevant quantity is the following
\begin{equation}
\label{20} \Delta A^2  :=  \bra{\Phi_N} \hat{A}^2 \ket{\Phi_N} -
\bra{\Phi_N} \hat{A} \ket{\Phi_N}^2 ,
\end{equation}
which identically vanishes if $\ket{\Phi_N}$ is an eigenstate of
$\hat{A}$. This quantity describes the quantum fluctuations of
observable $\hat{A}$ on state $\ket{\Phi_N}$. Since state
$\ket{\Phi_N}$ is a random variable, also $\Delta A^2$ will
fluctuate. Its average over the distribution of the coefficient
reads
\begin{equation}
\label{2} \delta_q A^2  := \overline{ \bra{\Phi_N} \hat{A}^2
\ket{\Phi_N}} - \overline{ \bra{\Phi_N} \hat{A} \ket{\Phi_N}^2}
\end{equation}
and involves the same quartic average that appears in (\ref{1}).
Being related to the average of $\Delta A^2$ in
Eq.\ (\ref{20}), $\delta_q A$ vanishes identically if the ensemble
is made up of eigenstates of $\hat{A}$. On the other hand, it may
also vanish asymptotically (for suitable values of $n$) as $N$
increases. If this happens, the outcome of a measurement of
observable $\hat{A}$ on the majority of states $\ket{\Phi_N}$ in
the ensemble is within good approximation fixed by
its expectation value $A$.

The key quantity is
\begin{eqnarray}
\label{deltahatA}
\delta A^2 & := & \delta_s A^2 + \delta_q A^2 \nonumber \\
& = & \overline{ \bra{\Phi_N} \hat{A}^2 \ket{\Phi_N}} - \overline{
\bra{\Phi_N} \hat{A} \ket{\Phi_N}}^2 .
\nonumber \\
& = & \tr (\rho_n \hat{A}^2 ) - {\!\left[ \tr (\rho_n \hat{A} )
\right]\!}^2.
\end{eqnarray}
A few comments are now in order. First of all, notice the
cancellation of the quartic terms and observe that this quantity
depends only on quadratic averages and is expressed in terms of
the density matrix (\ref{rhomix}) (as it should). As a matter of
fact, $\delta A^2$, namely the quantum variance of the observable
$\hat{A}$ on the microcanonical density matrix $\rho_n$, could
have been introduced without reference to $\delta_s A$ and
$\delta_q A$. The previous ``derivation'' aims only at elucidating
the multiple aspects of the fluctuations that affect a quantum
system in the framework we introduced. Since $\delta A^2$ controls both the
statistical variance $\delta_s A^2$ of the expectation value and
the average quantum variance $\delta_q A^2$ of the observable, the
condition
\begin{equation}
\label{deltaasym}
\frac{ \delta A}{ \overline{A} } \to 0
\end{equation}
ensures that, for the overwhelming majority of wave functions in
$\mathcal{H}_n$, an experimental measurement of the observable
$\hat{A}$ will fluctuate within a very narrow range around the
average expectation value $\overline{A}$. In conclusion, if
$\ket{\Phi_N} \in \mathcal{H}_n$, the outcome of a measurement of
$\hat{A}$ is with high accuracy \emph{independent of the experimental run}, i.e.\
independent of  the (in principle unknown) initial wave function. We call
this property \textit{typicality of the observable $A$}.
The asymptotic validity
of condition (\ref{deltaasym}) as $N\to\infty$ depends on the
choice of $\mathcal{H}_n$ and on how $n$ scales with $N$
\cite{typbec,FPPS}.

\section{Two-Mode Case Study}
\label{sec-casestudy}

Scrutiny of a simple case-study will hopefully elucidate the main
ideas and be the testbed of the general framework described in the
preceding section. Consider a two-mode system made up of $N$
structureless bosons. The second-quantized field operators satisfy
the canonical equal-time commutation relations (in this section we
remove the hats on the operators) \barr [\Psi(\bm{r}),
\Psi(\bm{r}')] = 0, \quad [\Psi(\bm{r}),  \Psi^{\dagger}(\bm{r}')]
= \delta(\bm{r}-\bm{r}'). \earr The $N$ bosonic particles are
distributed among the ground $\ket{\phi_0}$ and the first excited
state $\ket{\phi_1}$ of a harmonic oscillator, whose mode wave
functions read (in suitable units)
\begin{equation}
\phi_0(x) = \frac{1}{\pi^{1/4}} e^{-\frac{x^2}{2}}, \quad
\phi_1(x) = \frac{\sqrt{2}}{\pi^{1/4}} x e^{-\frac{x^2}{2}},
\end{equation}
and whose Hamiltonian is
\begin{equation}
H = \frac{1}{2} ( p^2 + x^2 ).
\end{equation}
One easily computes the expectation values of the even powers of
the position operator in the two modes
\begin{eqnarray}
\bra{\phi_0} x^{2\nu} \ket{\phi_0} & = & 2 \frac{(2\nu)!}{\nu!},
\\ \bra{\phi_1} x^{2\nu} \ket{\phi_1} & = & 2 \bra{\phi_0}
x^{2\nu+2} \ket{\phi_0} = 4 \frac{(2\nu+2)!}{(\nu+1)!},
\end{eqnarray}
while the expectation values of the odd powers vanish. Define the
collective single-particle observable
\begin{equation}
X_{2\nu} = \int dx\, x^{2\nu} \Psi^{\dagger}(x) \Psi(x).
\end{equation}
Let us consider the microcanonical ensemble represented by the
density matrix [see Eq.~(\ref{rhomix})]
\begin{equation}
\rho_n = \frac{1}{n} \sum_{| \ell| < n/2} \ket{\ell}
\bra{\ell},
\end{equation}
where in the states $\ket{\ell} := \ket{ (N/2+\ell)_0,
(N/2-\ell)_1 }$ the occupation numbers of the two modes are
well-defined, with $2\ell$ representing the particle imbalance
between the modes. One can easily show that, due to the symmetry
of the modes, this quantity is typical whenever the maximal
imbalance satisfies $n=o(N)$. The proof is simple. The expectation
value
\begin{eqnarray}
\overline{X_{2\nu}} &=& \mathrm{Tr} \!\left( \rho_n
X_{2\nu} \right)\! = \frac{N}{2} \!\left( \bra{\phi_0}
x^{2\nu} \ket{\phi_0} + \bra{\phi_1} x^{2\nu} \ket{\phi_1}
\right)
\nonumber\\
&=& N\left(\frac{(2\nu)!}{\nu!} + \frac{2 (2\nu+2)!}{(\nu+1)!}\right)
\end{eqnarray}
splits, as one expects, into the average of
expectation values of $x^{2\nu}$ in the two modes. Its variance on $\rho_n$ (\ref{deltahatA}) can be
expanded as a quadratic polynomial in the total number of particles $N$ and
the maximal imbalance $n$, yielding~\cite{FPPS}
\begin{eqnarray}
\delta X_{2\nu}^2 &=& \mathrm{Tr} \!\left(\rho_n X_{2\nu}^2
\right)\! - \!\left[
\mathrm{Tr} \!\left( \rho_n X_{2\nu} \right) \right]^2  \nonumber \\
&=& D_{2,0}^{(2\nu)}  \frac{N^2}{4}  +
D_{0,2}^{(2\nu)} n^2 + O(N).
\label{dx}
\end{eqnarray}
A straightforward computation shows that
\begin{equation}
D_{2,0}^{(2\nu)} = 2 \!\left| \bra{\phi_1} x^{2\nu} \ket{\phi_0}
\right|^2 = 0,
\end{equation}
due to the opposite symmetry of
the mode wave functions. By contrast, the factor multiplying $n^2$
does not vanish and reads
\begin{eqnarray}
D_{0,2}^{(2\nu)} &=& \frac{1}{12} \!\left( \bra{\phi_1} x^{2\nu}
\ket{\phi_1} - \bra{\phi_0} x^{2\nu} \ket{\phi_0} \right)^2
\nonumber\\
&=& \frac{1}{3}\left(\frac{2 (2\nu+2)!}{(\nu+1)!}-\frac{(2\nu)!}{\nu!} \right)^2.
\end{eqnarray}
These results show that, unless $n=O(N)$,
\begin{equation}
\frac{\delta X_{2\nu}}{ \overline{X_{2\nu}} } \to 0
\quad \textrm{as} \,\, N\to\infty ,
\end{equation}
thus ensuring the typicality of the observable $X_{2\nu}$ for
$n=o(N)$. In practice, no matters how one prepares the initial
state, distributing the particles between the two modes, as far as
the maximum imbalance between the two modes scales less fast than
the total number of particles $N$, a measurement of the collective
observable $X_{2\nu}$ will yield essentially the same result.
In particular from (\ref{dx}), when $n=O(N^{1/2})$, the relative fluctuations around the typical value are normal, i.e.\
$O(N^{-1/2})$.

\section{Conclusions and Outlook}
\label{sec-concl}

We have discussed the notion of quantum typicality, defining the
typicality of an observable and  focusing on a
two-mode Bose system. An observable is typical if its single-run
measurement, performed on a system state belonging to a suitable
subspace, yields the same result with very large probability.

Typical observables are therefore properties shared by the vast
majority of states. By contrast, non-typical observables are
characterized by wide fluctuations. Interestingly, this
distinction is crucial in determining ``good'' observables in
classical and quantum statistical mechanics \cite{landau}. As
measurements on typical observables yield (almost) the same
result, the knowledge of the initial state with
arbitrary precision becomes immaterial. This brings us back to the concepts
discussed in the Introduction and the main idea of this article.
One can revisit and relax the notions of state preparation and
initial conditions. As we emphasised, Bose-Einstein condensates
are an ideal testbed for these concepts in quantum statistical
physics \cite{PS,PeSm,Leggett,BDZ,Leggettbook}.

Typicality is related to the beautiful mathematical phenomenon of
measure concentration \cite{Ledoux}. This is a fecund idea that
has been applied to elucidate the structure of entanglement in
large quantum systems \cite{Winter,FMPPS}, as well as some basic
concepts in statistical mechanics
\cite{Tasaki,Popescu2,Goldstein,ref:Rigol-Nature}. It would
interesting to apply this notion to the characterization of
entanglement in Bose-Einstein condensates and to study dynamical
effects, such as phase randomization in condensates
\cite{Schmied5,FNPPSY}.

\section*{Acknowledgments}
We thank H.\ Nakazato and K.\ Yuasa for many interesting discussions.
This work was partially supported by Italian PRIN 2010LLKJBX on ``Collective quantum phenomena: from strongly correlated systems to quantum simulators'', and by the Italian National Group of Mathematical Physics (GNFM-INdAM).

\end{document}